# THE BARYONIC FRACTION IN GROUPS OF GALAXIES FROM X-RAY MEASUREMENTS

MARK J. HENRIKSEN[1] AND GARY A. MAMON[2]

## ABSTRACT

The recent *ROSAT* X-ray detections of hot intergalactic gas in three groups of galaxies are reviewed and the resulting baryonic fraction in these groups is reevaluated. We show that the baryonic fraction obtained, assuming hydrostatic equilibrium, should depend, perhaps sensitively, on the radius out to which the X-rays are detected, and the temperature profile of the gas. We find that the NGC 2300 group has a baryonic fraction out to 25' of at least 20%, thus over five times higher than in the original analysis of Mulchaey *et al.* (1993), and also much higher than one would obtain from big-bang nucleosynthesis, but similar to the other two groups as well as rich clusters. With this baryonic fraction, groups would be fair tracers of the distribution of baryons in the Universe if $\Omega h_{50}^2 = 0.3$. A baryonic fraction that increases with radius is consistent with the X-ray data from all three groups. However, a detailed analysis of the NGC 2300 group shows that the dependence of baryonic fraction on radius is not well constrained by the data, in part because of uncertainties in the estimated background.

*Subject headings:* galaxies: clusters of — galaxies: intergalactic medium — cosmology: observations —

## 1. INTRODUCTION

While clusters of galaxies commonly emit diffuse X-rays associated with a hot intergalactic medium, as evidenced by numerous X-ray observations with the *EINSTEIN* Observatory (*e.g.*, Forman & Jones 1982), the presence of detectable diffuse hot X-ray emitting gas in small groups of galaxies does not seem as common. *EINSTEIN* observations have revealed diffuse hot gas in four compact groups (Biermann, Kronberg & Madore 1982; Biermann & Kronberg 1983; Bahcall, Harris & Rood 1984), but failed to detect X-rays in 2 other compact groups, while a third compact group emitted X-rays that were probably associated with the individual galaxies (Bahcall *et al.*). Recent higher-sensitivity *ROSAT* observations with the Position Sensitive Proportional Counter (PSPC) have confirmed that some groups of galaxies emit X-rays originating from a hot diffuse intergalactic medium. Indeed, short (≃ 10 min) snapshots from the *ROSAT* survey mode has produced 12 X-ray detections among Hickson's (1982) 100 compact groups and these detected X-rays are probably of intergalactic origin in roughly 9 of these 12 compact groups (H. Ebeling, private communication). Deeper *ROSAT* PSPC images were recently obtained in four groups: the loose groups NGC 2300 (Mulchaey *et al.* 1993, hereafter MDMB), NGC 5044 (David *et al.* 1993), and the Hickson compact groups HCG 12 (H. Ebeling, private communication) and HCG 62 (Ponman & Bertram 1993). In only in one group (HCG 12) was there no intergalactic X-ray emission. The analysis of the X-ray properties of the NGC 2300 group has revealed a relatively low fraction of mass in gas and stars (hereafter, *baryonic fraction*). Mulchaey *et al.*'s analysis produced a baryonic fraction of 4%, and their upper limit was 15%. On the other hand, Ponman & Bertram's (1993) baryonic fraction was > 13%, while David *et al.* (1993) get 10%, both within 250 kpc (throughout this *Letter* we assume $H_0 = 50 \,\text{km}\,\text{s}^{-1}\,\text{Mpc}^{-1}$).

The baryonic fraction in clusters has been used as a constraint on the cosmological density parameter $\Omega$: if clusters are representative of the whole Universe, one has $\Omega = \Omega_b/f_b$ (White 1992). Standard big-bang nucleosynthesis produces a baryonic fraction on the scale of the whole Universe of $0.04 < \Omega_b h_{50}^2 < 0.08$ (Steigman 1989), where $h_{50} = H_0/(50 \,\text{km}\,\text{s}^{-1}\,\text{Mpc}^{-1})$. The baryonic fraction in clusters of galaxies is typically around 25% (*e.g.*, White 1992, and references therein). One thus obtains $\Omega \simeq 4\Omega_b \simeq 0.24$. As noted by MDMB, the implications of a baryonic fraction as small as $\Omega_b$ is that if the Universe has a density near closure ($\Omega_0 \simeq 1$), groups of galaxies would be fair tracers of the baryonic content of the Universe.

Another way to set constraints on $\Omega$ is to assume that groups are fair tracers of the ratio of mass to blue light in the Universe. The luminosity density of the Universe is such (Loveday *et al.* 1992) that the closure mass-to-light ratio is $M/L = 780\,h_{50}$, and in general one can write $\Omega = M/L/(780\,h_{50})$. The typical $M/L$ of groups (*e.g.*, Gourgoulhon *et al.* 1992), computed with the assumption that groups are in virial equilibrium, yields $\Omega = 0.07$, while a value of $\Omega = 0.3$ is obtained with the correction for the non-virialized cosmo-dynamical state of these systems (Mamon 1994).

In this *Letter*, we consider how the baryonic fraction in groups (and clusters) of galaxies should vary with radius, assuming simple models for the properties of the hot X-ray emitting diffuse gas. We analyze the published data for NGC 2300 using a similar model to that applied to rich clusters to reevaluate its baryonic fraction at the present limiting detection radius, and we indicate which future X-ray observations could place sufficient constraints on this parameter.

## 2. BARYONIC FRACTION VERSUS RADIUS

The X-ray surface brightness profile of groups and clusters are usually well-fitted by a law of the form

$$S(R) = S_0 \left[1 + (R^2/R_c^2)\right]^{1/2 - 3\beta} \qquad (1)$$

unless these systems possess a central cooling flow, which

[1] Dept. of Physics and Astronomy, University of Alabama, Tuscaloosa AL 35487, USA
[2] DAEC (also CNRS URA 173 and associated with Université Denis Diderot [Paris 7]), Observatoire de Paris-Meudon, F-92195 Meudon, FRANCE

produces a central peak in the surface brightness profile. Here $\beta$ is an empirical rather than a physical parameter

Inverting the Abel equation relating X-ray surface brightness to the 3D emissivity in the spectral passband of the instrument, noting that the emissivity of a hot plasma varies as $n^2 \Lambda(T)$, where $\Lambda$ is the cooling function, calling $\eta = d \ln \Lambda / d \ln T$, and assuming an equation of state $T \sim n^{\gamma-1}$ (i.e., isothermal for $\gamma = 1$ and polytropic otherwise), one obtains (see Cowie, Henriksen & Mushotzky 1987)

$$n(r) = n_0 \left[1 + (r^2/R_c^2)\right]^{-\delta} \quad (2a)$$

$$T(r) = T_0 \left[1 + (r^2/R_c^2)\right]^{-\delta(\gamma-1)} \quad (2b)$$

$$\delta = \frac{3\beta/2}{1 + \eta(\gamma-1)/2} \quad (2c)$$

From equation (2a), the total mass in gas can be written

$$M_{\rm gas}(r) = 4\pi n_0 R_c^3 \mu m_p \tilde{M}(x) , \quad (3a)$$

$$x = r/R_c , \quad (3b)$$

$$\tilde{M}(x) = \int_0^x \frac{y^2 \, dy}{(1+y^2)^\delta} , \quad (3c)$$

where $\mu$ is the mean particle weight in units of the proton mass $m_p$.

$$\tilde{M}(x) = \frac{x}{2}\left(x^2+1\right)^{1/2} - \frac{1}{2}\sinh^{-1} x \quad \text{for } \delta = 1/2 , (4a)$$

$$\tilde{M}(x) = x - \tan^{-1} x \quad \text{for } \delta = 1 , \quad (4b)$$

$$\tilde{M}(x) = \sinh^{-1} x - \frac{x}{(x^2+1)^{1/2}} \quad \text{for } \delta = 3/2 . (4c)$$

Writing the equation of hydrostatic equilibrium, the total mass (that binds the gas) is (Cowie, Henriksen & Mushotzky 1987)

$$M_{\rm tot} = -\frac{kTr}{G\mu m_p}\left(\frac{d\ln n}{d\ln r} + \frac{d\ln T}{d\ln r}\right)$$

$$= 2\gamma\delta \frac{kT_0 R_c}{G\mu m_p} \frac{x^3}{(1+x^2)^{1+\delta(\gamma-1)}} . \quad (5)$$

Using equations (3) and (5), the gas fraction can be written

$$f_g = \frac{2\pi G \mu^2 m_p^2 n_0 R_c^2}{\gamma \delta k T_0}(1+x^2)^{1+\delta(\gamma-1)}\frac{\tilde{M}(x)}{x^3} . \quad (6)$$

Figure 1 shows the gas fraction versus radius for isothermal models with different $\delta$s, scaled to $T = 1\,{\rm keV}$, $R_c = 100\,{\rm kpc}$, and $n_0 = 10^{-3}\,{\rm cm}^{-3}$. In general, the gas fraction should be multiplied by

$$(n_0/10^{-3}\,{\rm cm}^{-3})(R_c/100\,{\rm kpc})^2(T_0/1\,{\rm keV})^{-1} . \quad (7)$$

From Figure 1, one sees that for $\delta = 1/2$, the gas fraction increases sharply with radius, while for $\delta = 3/2$, the reverse is true, and finally for $\delta = 1$ the gas fraction is asymptotically constant, and equal to 2.3 times its core radius value. From equation (6), the gas fraction varies asymptotically as

$$f_g \sim (r/R_c)^{2-\delta[2-(\gamma-1)]} \quad (8)$$

so that a constant asymptotic gas fraction implies $\delta = 1$ for isothermal models and $\delta = 3/2$ for $\gamma = 5/3$ polytropes.

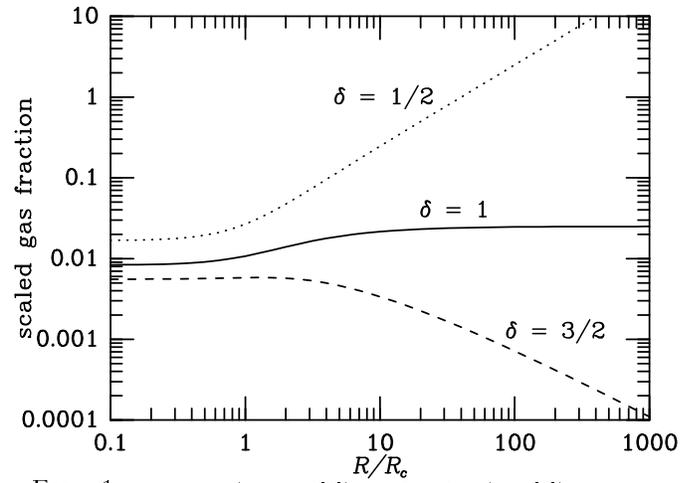

FIG. 1.— Scaled (see eq. [7]) gas fraction (eq. [6]) vs. radius for isothermal models.

Figure 2 shows the same as Figure 1, but for polytropic models (all with $\delta = 1$). Again the gas fraction should in general be multiplied as in equation (7). For non-isothermal gas distributions, the gas fraction increases with radius outside of the central region.

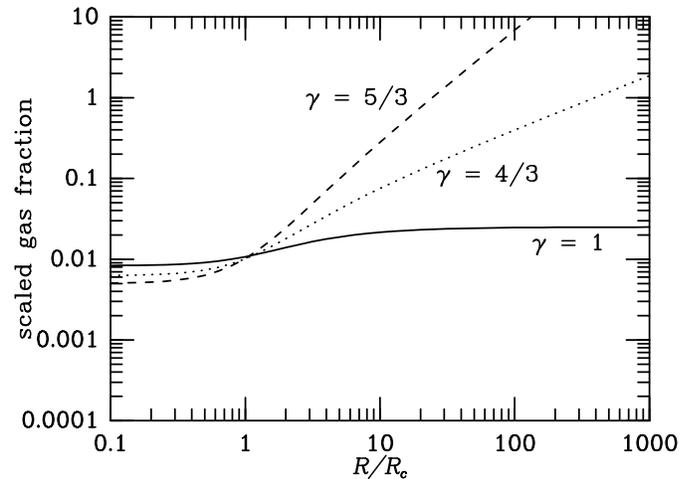

FIG. 2.— Scaled (see eq. [7]) gas fraction (eq. [6]) vs. radius for $\delta = 1$ polytropic models.

Note that certain parameters ($\delta = 1/2$ with $\gamma = 1$ or $\delta = 1$ with $\gamma = 5/3$) yield unphysical gas fractions over unity at large radii ($R \gtrsim 100 R_c$).

## 3. COMPARISON WITH OBSERVATIONS

The *ROSAT* X-ray observations are not strongly constraining for NGC 2300. For example, in Figure 3, we plot the intergalactic X-ray surface brightness profile of the NGC 2300 group. The points are from MDMB (where we omit the three points at $18'$, $21'$ and $24'$ for which there is obstruction from the window support structure) and uncorrected for the background while the curves are fits to these points using the model in equation (1), with a constant background $S_{bg}$ added everywhere. Of course, MDMB have determined the background independently of their NGC 2300 observations, but their adopted value of $S_{bg} = 7.0 \times 10^{-4}\,\mathrm{arcmin}^{-2}\,\mathrm{s}^{-1}$ (D. Burstein, private communication), could be uncertain by 10% or more, and this is why we allow it to be a free parameter. Figure 3 clearly shows that very different good fits (without the point at $39'$, which is well above the surrounding points and may be contaminated by a point source) provide very different $\delta$ and core radii.

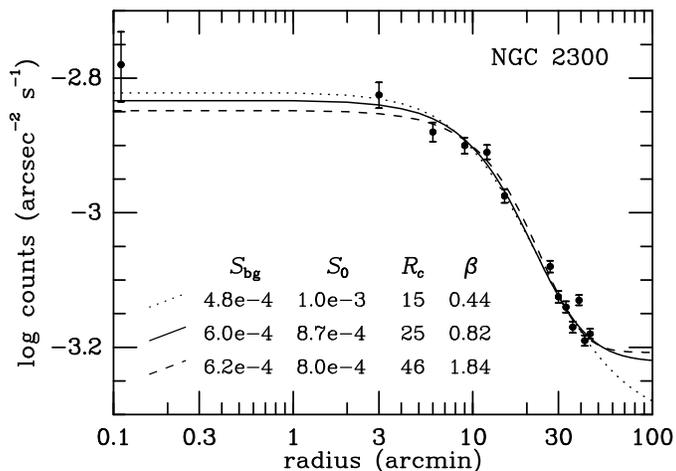

FIG. 3.— Surface brightness profile of the NGC 2300 group intergalactic X-ray emission. Data (uncorrected for background) from Mulchaey *et al.* (1993) are shown as *points*, while the best fits to isothermal-$\beta$ models (eq. [1] plus a constant background) shown in the top three lines of Table 1 are respectively shown as *dotted*, *solid*, and *dashed curves*. Points at $18'$, $21'$, and $24'$ are omitted as they suffer from obstruction from the window support structure of the PSPC.

Using the results from §II, the top three rows in Table 1 illustrate how these three fits return different total mass ($M_{\rm tot}$, eq. [5]), gas mass ($M_{\rm gas}$, eqs. [3] and [4]), gas fraction ($f_g$, eqs. [4] and [6]), and baryonic fraction ($f_b$) within $25'$ of the center of the diffuse gas in the NGC 2300 group. The fourth row in Table 1 lists what would be obtained if we adopted the MDMB parameters of $S_{bg} = 7.0 \times 10^{-4}$, $\beta = 0.85$ (J. Mulchaey, private communication), and in subsequent rows, we investigate alternative fits imposing either $R_c$ or $\beta$ in addition to $S_{bg}$. As MDMB, we assume isothermality and we convert gas fractions to baryonic fractions using $M_{\rm gal}(25') = 6 \times 10^{11} h_{50}^{-1} M_\odot$. The gas number density necessary to normalize the gas fractions (eq. [3a]) is obtained by writing the bolometric luminosity as

$$L_{\rm bol} = \int n^2 \Lambda(T) dV = 4\pi n_0^2 R_c^3 \Lambda(T_0) \int_0^{R_{\rm max}/R_c} x^2/(1+x^2)^{2\delta} dx ,$$

where $\Lambda$ is the Raymond and Smith (1977) cooling function. The density calculation uses the cooling function evaluated for the emission weighted temperature (0.86 keV) and is $8.6 \times 10^{-24}\,\mathrm{ergs\,cm^3\,s^{-1}}$ (using a heavy element abundance of 6%, the best fit value of MDMB) and $R_{\rm max} = 25'$ is the extent of the X-ray emitting hot gas.

TABLE 1
REANALYSIS OF THE NGC 2300 GROUP FOR $R < 25'$

| $S_{bg}$ | $S_0$ | $R_c$ | $\beta$ | $\chi^2$ | $L_{\rm bol}$ | $n_0$ | $M_{\rm gas}$ | $M_{\rm tot}$ | $f_g$ | $f_b$ |
| (1) | (2) | (3) | (4) | (5) | (6) | (7) | (8) | (9) | (10) | (11) |
|---|---|---|---|---|---|---|---|---|---|---|
| 4.8 | 10.3 | 15.0 | 0.44 | 19.9 | 49.0 | 1.97 | 2.49 | 10.2 | 0.24 | 0.30 |
| 6.0 | 8.7  | 25.0 | 0.82 | 18.8 | 38.7 | 1.66 | 2.21 | 13.0 | 0.17 | 0.22 |
| 6.2 | 8.0  | 46.0 | 1.84 | 16.8 | 38.0 | 1.48 | 2.21 | 13.3 | 0.17 | 0.21 |
| 7.0 | 10.5 | 15.0 | 0.85 | 148.3 | 23.8 | 2.14 | 1.61 | 19.9 | 0.08 | 0.11 |
| 7.0 | 8.9  | 18.0 | 0.85 | 121.2 | 26.4 | 1.84 | 1.76 | 17.7 | 0.10 | 0.13 |
| 7.0 | 9.7  | 15.0 | 0.76 | 138.1 | 25.3 | 2.04 | 1.70 | 17.9 | 0.09 | 0.13 |
| 6.5 | 9.8  | 15.0 | 0.65 | 66.6 | 30.9 | 1.99 | 1.92 | 15.1 | 0.13 | 0.17 |
| 6.5 | 8.7  | 25.0 | 0.99 | 35.8 | 33.0 | 1.68 | 2.01 | 15.7 | 0.13 | 0.17 |

NOTES: (1) Background in $10^{-4}\,\mathrm{arcmin}^{-2}\,\mathrm{s}^{-1}$. (2) Central surface brightness in $10^{-4}\,\mathrm{arcmin}^{-2}\,\mathrm{s}^{-1}$. (3) Core radius in arcmin. (4) Surface brightness shape (eq. [1]). (5) Goodness of fit. (6) Bolometric luminosity in $10^{42}\,\mathrm{ergs\,s}^{-1}$. (7) Central gas density in $10^{-3}\,\mathrm{cm}^{-3}$. (8) Gas mass within $25'$ in $10^{12} M_\odot$. (9) Total mass within $25'$ in $10^{12} M_\odot$. (10) Gas fraction within $25'$. (11) Baryonic fraction within $25'$.

Table 1 shows that the best fits (rows 1 to 3) conspire to a surprisingly robust gas mass within $25'$. Moreover, the fit with the MDMB parameters is significantly worse than the three fits shown in Figure 3, because their adopted background is too high. The best fit baryonic fractions are $\gtrsim 20\%$, over 5 times larger than the value quoted by MDMB (4%). In fact, our analysis shows that even the MDMB parameters should return a baryonic fraction as high as 11%. The discrepancy with MDMB's low result is due in part to MDMB probably computing the baryonic fraction within $15'$ (as stated in the main body of their paper) while in their *Note added in proof*, they mention a total mass one-third lower within a region extending two-thirds further ($25'$). For comparison, NGC 5044 has a baryonic fraction of $\simeq 10\%$ within 250 kpc (David *et al.* 1993) and HCG 62 $> 13\%$ within a similar region. Also, David *et al.* detect diffuse X-ray emission out to 400 kpc. Since NGC 2300 fills the PSPC field of view, the extent of its gas is $> 0.5$ Mpc.

The first two groups have low values of $\delta$: Ponman's & Bertram's analysis of HCG62 yields $\delta = 0.54$, while David *et al.* find $\delta = 0.79$ for NGC 5044. Also, the temperature profiles for NGC 5044 and HCG 62 are slightly non-isothermal outside of the cooling radius with effective $\gamma$ of 1.1 and 1.15 respectively. These values of $\delta$ and $\gamma$ imply that the gas fraction in HCG62 and NGC 5044 is increasing with radius near the limit of the *ROSAT* detections. On the other hand, the NGC 2300 data is not constraining enough to evaluate $\delta$ (from eq. [2c], $\delta = 3\beta/2$ for the assumed isothermal model). Indeed, decent fits yield $\delta = 0.65$ to $\delta > 2.5$. The high background used by MDMB forces a high asymptotic slope for the surface brightness profile, yielding values of $\delta$ much higher than that found for the other groups and for clusters.

## 4. DISCUSSION

Groups are probably more representative of the Universe than are clusters, as they include altogether perhaps five times as many galaxies as do clusters. With a baryonic fraction over 20%, the inner $25'$ ($330\,h_{50}^{-1}$kpc) of the NGC 2300 group would be a fair tracer of the baryonic fraction in the Universe if $\Omega \simeq 5\,\Omega_b \simeq 0.3 h_{50}^{-2}$. These numbers are similar to the constraints from group $M/L$s, with (if $H_0 = 50$) and without (if $H_0 = 100$) corrections for non-virialized states, respectively (see §I).

From equation (8), a constant baryonic fraction can be reached with $\delta = 1$ isothermal gas or with $\delta = 3/2$ polytropic gas with index $\gamma = 5/3$. Figure 1 shows that for isothermal $\delta = 1$ gas, the asymptotic gas fraction is roughly 2 to 2.5 times larger than at $25'$. Therefore, an extrapolation of the gas fraction to large scales will yield a baryonic fraction $f_b \geq 40\%$, and will thus be consistent with baryonic nucleosynthesis if $\Omega \leq 0.15\,h_{50}^{-2}$. Taken at face value, the observations of the three groups discussed here suggest $\delta < 1$ and $\gamma > 1$, both of which imply baryonic fractions that *increase* with radius, which extrapolates at large scales to even lower values of $\Omega$. However, as noted in §III, high values of $\delta$ can produce equally good fits to the NGC 2300 data, and thus produce constant or decreasing baryonic fractions. Note that a similar trend of increasing gas fraction is found in clusters (*e.g.*, Durret *et al.* 1993).

To reconcile the data for groups with $\Omega = 1$, as favored by inflation, one requires a baryonic fraction that *decreases* at large radii to an asymptotic value near $\Omega_b \simeq 0.06\,h_{50}^{-2}$, hence implying $\delta > 1$ (isothermal gas) or $\delta > 3/2$ ($\gamma = 5/3$ polytropic gas). Although these values of $\delta$ seem inconsistent with the X-ray observational data of the three groups studied here, there are two possibilities to remain consistent with inflation: 1) While the gas fraction is nearly constant within the group, it could fall to zero outside of the group, so that groups of galaxies would be sites of higher baryonic fraction in the Universe, *i.e. biased* tracers of the distribution of the baryonic fraction in the Universe; 2) The cosmological constant, $\Lambda$ (not to confuse with the cooling function used above with the same symbol) is non-zero, *i.e.*, $\Omega = 0.2$ and $\Lambda = 0.8$, which is consistent with the distribution of cluster temperatures deduced from X-ray observations (Bartlett & Silk 1993).

The model surface brightness profiles of the NGC 2300 group have a similar shape in the inner region, and are thus indistinguishable, although uncertainties in the PSPC background subtraction allow these profiles to diverge significantly beyond $R > 45'$, as seen in Figure 3. This uncertain behavior at large radii for NGC 2300 and the increasing baryonic fraction in HCG62 and NGC 5044 make clear that it is of fundamental importance to observe X-ray emission at relatively large distances from the centers of groups to determine the extent of the gas.

As a final note, if the three groups studied here are still in the stages of cosmological collapse (as is argued for *all* non-compact groups by Diaferio *et al.* 1993 and Mamon 1994), then the equation of hydrostatic equilibrium used here (eq. [5]) may not apply, for two reasons: 1) The gas may not have time to react to the rapid changes of the global potential of the group; 2) Even if the gas follows the potential of the whole group, one needs to add a term $\partial(\rho \bar{v}_r)/\partial t$ in equation (5). Ponman & Betram (1993) conclude that the compact group HCG 62 is past full collapse and currently in the process of slowly shrinking by orbital energy dissipation via dynamical friction. In general, it is reasonable to assume that during the collapse of a group, the denser inner parts will collapse earlier and be close to equilibrium today, the question then being how far out is the gas in equilibrium. It may well be that if hot intergalactic gas found in a group presents unusual properties, this could reflect a departure from hydrostatic equilibrium possibly caused by the group's cosmological collapse.

We thank Dave Burstein, Harald Ebeling, John Mulchaey, Jack Sulentic and Joe Silk for useful discussions, as well as Renato Dupke, Jack Sulentic and an anonymous referee for detailed and useful comments on the manuscript. This research was partly sponsored through NSF EPSCoR grant 8DP–UA2–92 and a travel grant from CNFA.